\documentclass[8.5pt,twoside,twocolumn]{article}
\usepackage[super,sort&compress,comma]{natbib}
\usepackage[version=3]{mhchem}
\usepackage{times}
\usepackage{sectsty}
\usepackage{balance}

\usepackage{graphicx}
\usepackage{lastpage}
\usepackage[format=plain,singlelinecheck=false,font=small,labelfont=bf,labelsep=space]{caption}

\newcommand{\degree}{\ensuremath{^\circ}}
\usepackage{textcomp}

\begin{document}

\thispagestyle{plain}
\renewcommand{\thefootnote}{\fnsymbol{footnote}}
\renewcommand\footnoterule{\vspace*{1pt}
\hrule width 3.4in height 0.4pt \vspace*{5pt}}
\setcounter{secnumdepth}{5}

\makeatletter
\def\subsubsection{\@startsection{subsubsection}{3}{10pt}{-1.25ex plus -1ex minus -.1ex}{0ex plus 0ex}{\normalsize\bf}}
\def\paragraph{\@startsection{paragraph}{4}{10pt}{-1.25ex plus -1ex minus -.1ex}{0ex plus 0ex}{\normalsize\textit}}
\renewcommand\@biblabel[1]{#1}
\renewcommand\@makefntext[1]
{\noindent\makebox[0pt][r]{\@thefnmark\,}#1}
\makeatother
\renewcommand{\figurename}{\small{Fig.}~}
\sectionfont{\large}
\subsectionfont{\normalsize}

\setlength{\columnsep}{6.5mm}
\setlength\bibsep{1pt}

\twocolumn[
  \begin{@twocolumnfalse}
\noindent\LARGE{\textbf{Rate-dependent elastic hysteresis during the peeling of Pressure Sensitive Adhesives}}
\vspace{0.6cm}

\noindent\large{\textbf{Richard Villey,\textit{$^{a,b}$}  Costantino Creton,\textit{$^{a}$}  Pierre-Philippe Cortet,\textit{$^{b}$} Marie-Julie Dalbe,\textit{$^{c,d}$} Thomas Jet,\textit{$^{a}$} Baudouin Saintyves,\textit{$^{a,b}$} St\'ephane Santucci,\textit{$^{d}$} Lo\"{i}c Vanel,\textit{$^{c}$}   David J. Yarusso\textit{$^{e}$} and Matteo Ciccotti \textit{$^{a\ast}$}}}\vspace{0.5cm}

\vspace{0.6cm}

\noindent
\normalsize{The modelling of the adherence energy during peeling  of Pressure Sensitive Adhesives (PSA)  has received much attention since the 1950's, uncovering several factors that aim at explaining their high adherence on most substrates, such as the softness and strong viscoelastic behaviour of the adhesive, the low thickness of the adhesive layer and its confinement by a rigid backing. The more recent investigation of adhesives by probe-tack methods also revealed the importance of cavitation and stringing mechanisms during debonding, underlining  the influence of large deformations and of the related non-linear response of the material, which also intervenes during peeling. Although a global modelling of the complex coupling of all these ingredients remains a formidable issue, we report here some key experiments and modelling arguments that should constitute an important step forward. We first measure a non-trivial dependence of the adherence  energy on the loading geometry, namely through the influence of the peeling angle, which is found to be separable from the peeling velocity dependence. This is the first time to our knowledge that such adherence energy dependence on the peeling angle is systematically investigated and unambiguously  demonstrated. Secondly, we reveal an independent strong influence of the large strain rheology of the adhesives on the adherence energy. We complete both measurements with a microscopic investigation of the debonding region.
We discuss existing modellings in light of these measurements and of recent soft material mechanics arguments, to show that the adherence energy during peeling of PSA should not be associated to the propagation of an interfacial stress singularity. The relevant deformation mechanisms are actually located over the whole adhesive thickness, and the adherence energy during peeling of PSA should rather be associated to the energy loss by viscous friction and by rate-dependent elastic hysteresis.}

\vspace{0.5cm}
 \end{@twocolumnfalse}
  ]

\section{Introduction and motivations}

\footnotetext{\textit{$^{a}$~Laboratoire SIMM, UMR7615 ESPCI-CNRS-UPMC-PSL,  France.}}
\footnotetext{\textit{E-mail: matteo.ciccotti@espci.fr}}
\footnotetext{\textit{$^{b}$~Laboratoire FAST, UMR7608 CNRS, Univ. Paris-Sud, France.}}
\footnotetext{\textit{$^{c}$~Institut Lumi\`ere Mati\`ere, UMR5306 Universit\'e Lyon 1-CNRS,Universit\'e de Lyon, France.}}
\footnotetext{\textit{$^{d}$~Laboratoire de Physique de l'ENS de Lyon, UMR5672 CNRS and Universit\'e de Lyon, France. }}
\footnotetext{\textit{$^{e}$~3M Center, 3M Company, 230-1D-15, St. Paul, MN, 55144-1000, USA. }}

\begin{figure}[t]
\centering
\includegraphics[width=8cm]{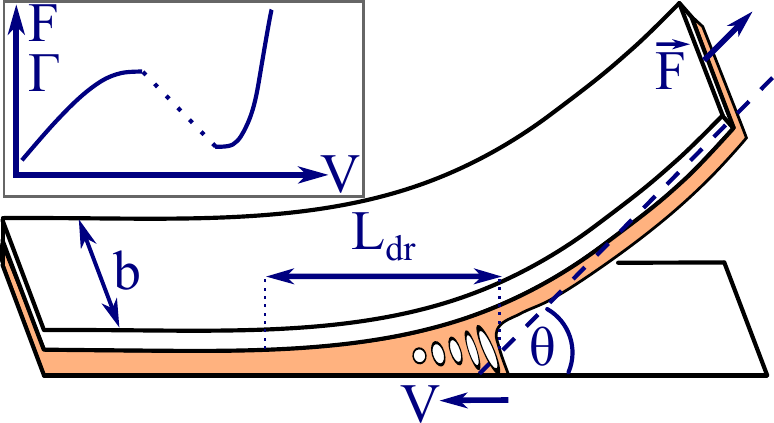}
 \caption{Geometry of a peeling experiment and typical variations of the peeling force $F$ and of the adherence energy $\Gamma$ with the peeling velocity $V$. $L_{dr}$ is the characteristic extension of the debonding region, where the adhesive is significantly strained. }\label{fig1}
\end{figure}

During the peeling of a Pressure Sensitive Adhesive (PSA), the adherence energy $\Gamma$ (the work which should be provided to peel a unit tape area) is several orders of magnitude above the thermodynamic Dupr\'e surface energy $w$ between the adhesive and the underlying substrate. This demonstrates the dominant role of energy dissipation.
Peeling can occur through failure inside of the adhesive layer (``cohesive failure") or through debonding of the adhesive from the substrate (``interfacial" or ``adhesive" failure).\cite{Gent1969} The latter is the most typical and useful failure mode for PSA, since it leaves the substrate clean: we  thus focus on interfacial failure in this paper.
Moreover, the adherence energy $\Gamma$ has a strong dependence on the peeling velocity  $V$ (see the insert in Fig.\ \ref{fig1}), which   presents a time-temperature equivalence with shift factors similar to those of the linear rheology of the adhesive. \cite{Kaelble1964,Gent1969,Maugis1985} This has suggested for a long time that small strain viscoelasticity is mainly responsible for the dissipation energy $\Gamma$, leading to two main modelling strategies.

The first approach \cite{Kaelble1964,Gent1969,Hata1972,Derail1998,Yarusso1999}  relates back to the 1960's and treats the adhesive layer as a (visco)elastic foundation, made of a parallel array of springs (and dashpots) linking the flexible tape backing to the underlying substrate. This foundation-based approach is equivalent to treating the adhesive layer as a cohesive zone linking two interfaces (the substrate and the backing).\cite{Dugdale1960,Barenblatt1962,Needleman1987}
Energy dissipation occurs in the whole thickness and width of the adhesive layer and affects a stress concentration region close to the peeling front. The extension $L_{dr}$ of this debonding region (see Fig.\ \ref{fig1}) is determined by the scale over which stress is transferred from the tape backing to the adhesive layer. $L_{dr}$  is typically several times the thickness of this adhesive layer.\cite{Kaelble1960} The link with rheology is made through the time scale $t^*$ associated to the strain  rate $\dot\varepsilon$ of the adhesive in this region, caused by the propagation of the peeling front at velocity $V$, namely $t^*\sim 1/\dot \varepsilon \sim L_{dr}/(V\varepsilon_{max})$, where $\varepsilon_{max}$ is the maximum stretch experienced by the adhesive (at the peeling front). In this theoretical framework, the velocity at which the characteristic maximum of adherence energy $\Gamma(V)$ is observed  is often attributed to the onset of the glass transition stiffening of the adhesive. The critical velocity associated to this  maximum also determines the onset of the stick-slip instability of the peeling (independently of the chosen model).

In the second approach, \cite{Schapery1975,DEGENNES1988,Saulnier2004,Persson2005,Barthel2009} dissipation is modelled as a viscoelastic perturbation of the inverse square root stress singularity of Linear Elastic Fracture Mechanics (LEFM). This singularity propagates at the interface between the adhesive and the substrate. Energy dissipation takes place in a region neighbouring the crack tip where the local strain rate (associated with the crack front propagation velocity $V$) corresponds to the viscoelastic relaxation time range of the adhesive. In this model, the peak in $\Gamma(V)$ is obtained when the size of this dissipative region becomes comparable to the adhesive thickness. \cite{DEGENNES1988}

In this singularity-based approach, the peeling energy $\Gamma(V)$ is interpreted as an interfacial fracture energy amplified by viscoelasticity and should therefore be independent of the geometry and of the loading conditions of the adhesive joint. On the contrary,  $\Gamma(V)$ in the other (the foundation-based) approach is associated to deformations acting over the whole adhesive thickness: the propagation of the interfacial  crack tip singularity plays a minor role and is not explicitly accounted for in the modellings. For this reason,  the measured adherence energy $\Gamma$  should be more properly interpreted as a work of debonding in this foundation-based  approach: it is only an apparent fracture energy since it is not a fundamental property of the interface between the adhesive and the substrate.

>From an experimental point of view, the $\Gamma(V)$ curves of soft confined adhesives were shown to be dependent on the adhesive thickness.\cite{Kaelble1992} An additional dependence of $\Gamma$ on the peeling angle $\theta$  can  be inferred from data available in the literature (see for example Figs 8-11 in Ref. \cite{Kaelble1960}). Nevertheless, no clear and direct demonstration of the $\Gamma(\theta)$ variations  is reported, and a systematic experimental investigation of this dependence has never been conducted. The fact $\Gamma$ depends on these two parameters tends to be in favour of the foundation-based  approach.  Data in the literature are however related to a lot of different types of adhesives, with a large diversity of liquid/solid behaviours. When considering soft solids only (such as most commercial PSA), this foundation-based  approach, especially Kaelble's model, \cite{Kaelble1960,Kaelble1964} seems to describe quite well the peeling experiments, as long as subtle choices are made regarding the model  parameters (\textit{e.g.}, the adhesive Young's modulus or the critical stress $\sigma_c$ at debonding). Kaelble's model is essentially linear and elastic, since viscoelasticity is only included through the change of the storage modulus with the characteristic time scale $t^*$. This model has however an unclear mechanical foundation: it would lead to a large and geometry dependent energy dissipation even in a purely Hookean material. This is in apparent contradiction with the energy analysis of Griffith on hard solids, extended to soft solids by Rivlin and Thomas:\cite{Rivlin1953} this analysis treats the adherence energy $\Gamma$ as an interfacial fracture energy, which should be a characteristic property of the interface between the adhesive and the substrate, and thus be independent of the geometry of the adhesive joint as well as on the loading conditions.

Furthermore, while several authors acknowledge the presence of long fibrils in the debonding region,\cite{Kaelble1965,Niesiolowski1981,Schapery1975,Persson2005} these are not explicitly included in their modellings.  The presence of fibrils however clearly suggests that the large strain mechanics and non-linear rheology of the adhesive play an important role in setting the adherence energy, as suggested by Gent and Petrich.\cite{Gent1969}

The aim of this paper is to examine the physics of these different modellings in light of recent developments in soft materials mechanics and large strain rheology. The strategy we follow has four steps:
\begin{enumerate}
\item
In order to understand the coupling between geometry, loading and  dissipation, we perform peeling experiments on a well-known commercial PSA,  in which the effects of the peeling angle $\theta$ and of the peeling velocity $V$ on the adherence energy $\Gamma$ are  systematically studied in an independent manner.
\item
We examine the impact of large strain rheology on  $\Gamma$ by performing peeling experiments on a series of custom-made PSA, for which linear and non-linear rheology are modified as independently as possible, yet remaining close to the rheology of  commercial PSA.
\item
During all these peeling experiments, we perform microscopic visualizations of the debonding region, in order to monitor the size and shape of  the fibrillated domain.
\item
Using soft mechanics arguments, we justify that the foundation-based   approach detailed above is the most relevant, and we test the ability  of different models within this category to describe our experimental results. We finally propose some key ingredients that should guide the development of a thorough  modelling, able to capture all the subtle debonding mechanisms for PSA based on soft and confined viscoelastic solids.
\end{enumerate}

\section{Materials and Methods}

Systematic peeling experiments at room temperature ($23\pm2\degree$C) are conducted on seven different types of
PSA. A commercial tape (3M Scotch\textregistered~600) is first used to investigate the influence of the peeling angle $\theta$ and peeling velocity $V$  on the  adherence energy $\Gamma$. This adhesive has been frequently used in the literature, including most of our previous investigations, \cite{Barquins1997,Amouroux2001,Cortet2007,Cortet2013,Dalbe2014,Dalbe2014d} mainly because of its quality and robustness, leading to very reproducible peeling experiments. We examine peeling angles  between 30 and 150$\degree$, to cover a broad range  while avoiding the very small or very large peeling angles, where unwanted processes can become dominant, such as plastic deformation of the tape backing \cite{Gent1977,Derail1997} or failure due to shear or slippage.\cite{Kaelble1960,Newby1997,Amouroux2001}

\begin{table}[bp]
\small
\centering
\caption{\ Compositions of the six custom-made copolymers used for the adhesives. EHA stands for 2-ethyl hexyl acrylate, MA for methyl acrylate and AA for acrylic acid.}
\label{Table1}
\begin{tabular*}{0.48\textwidth}{@{\extracolsep{\fill}}cccccc}
    \hline
    Name  & EHA & MA & AA & Cross-Linker & $T_g$ \\
    \hline
    1A & 70\% & 25\% & 5\% & 0.2\% & $-34\pm4$ $\degree$C  \\
    1B & 70\% & 25\% & 5\% & 0.4\% & $-34\pm4$ $\degree$C   \\
    2A & 85\% & 10\% & 5\% & 0.2\% & $-43\pm5$ $\degree$C   \\
    2B & 85\% & 10\% & 5\% & 0.4\% & $-43\pm5$ $\degree$C   \\
    3A & 95\% & 0\% & 5\% & 0.2\% & $-54\pm8$ $\degree$C   \\
    3B & 95\% & 0\% & 5\% & 0.4\% & $-54\pm8$ $\degree$C   \\
    \hline
\end{tabular*}
\end{table}

Since the second part of our study requires variations in the rheological properties of the adhesive  material, we use six different custom-made adhesive tapes, synthesized and coated in the 3M Research Center, in order to modify both the linear and non-linear rheologies as independently as possible. Each material is synthesized in solution and composed of various proportions of the monomers 2-ethylhexyl acrylate (EHA), methyl acrylate (MA) and acrylic acid (AA). Detailed compositions are given in table \ref{Table1}. Each adhesive solution is then coated on a PET film (thickness $2h=38$~\textmu m) in order to obtain a dry thickness of the adhesive of $a=20$~\textmu m. Before coating, 0.2 wt\% or 0.4 wt\% of aluminium acetyl acetonate (cross-linker) is added to each of the three adhesive compositions (labelled 1, 2, 3) in order to provide two levels of cross-linking (labelled A, B) to the adhesive film during drying. Acetyl acetate is also used as an inhibitor to have a better control on the cross-linking process.

\begin{figure}[btp]
\centering
\includegraphics[width=8.6cm]{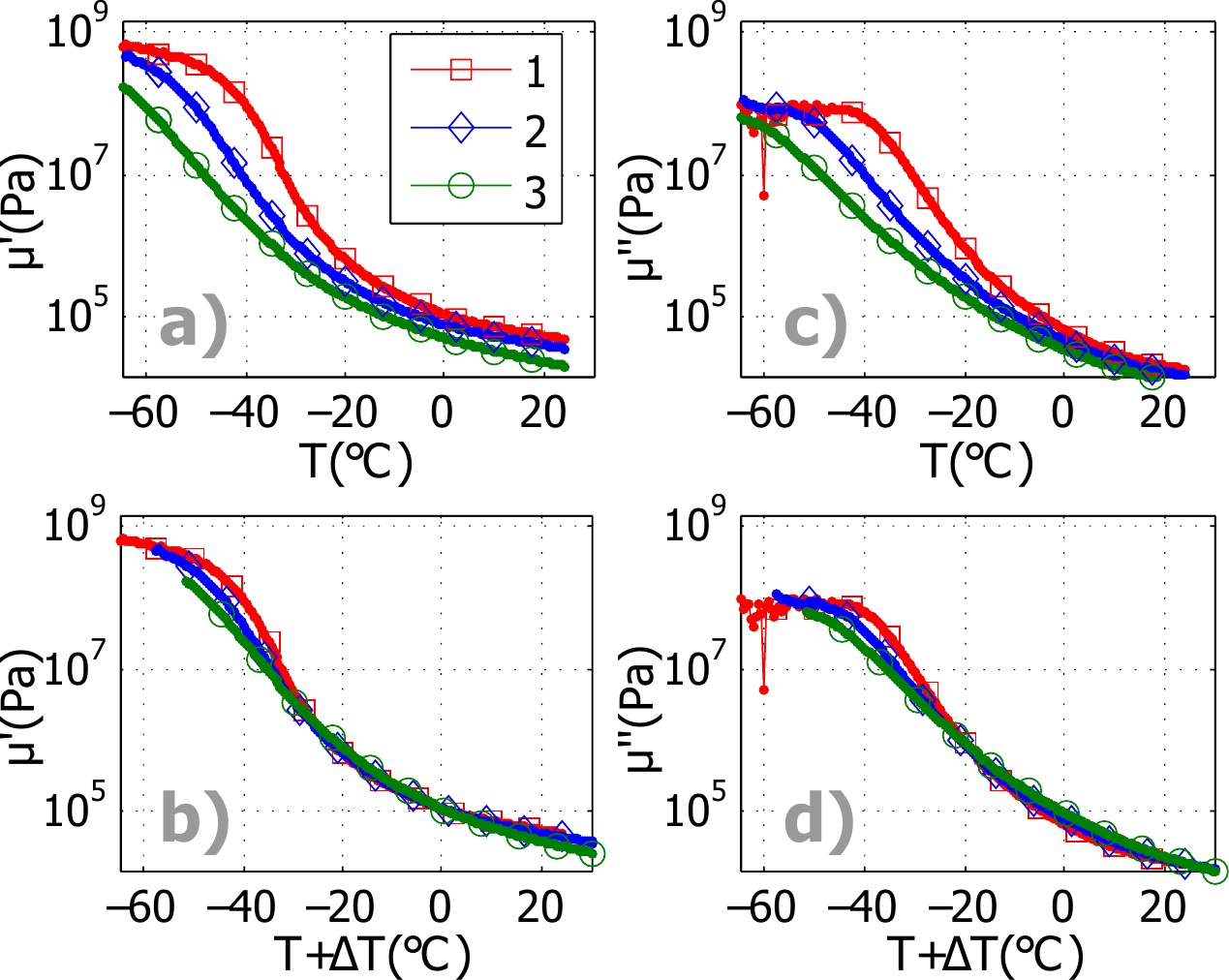}
\caption{(a) and (c): Storage $\mu'$ and loss $\mu''$ shear moduli of the uncross-linked adhesives from table \ref{Table1}, acquired at $\omega=1$ rad/s with strains smaller than $1\%$.
(b) and (d): Master curves obtained by shifting the shear moduli of the adhesives \#2 and \#3 by 6.5 and 13 \degree C respectively, in order to collapse the low modulus domain (below 1 MPa), as discussed in section \ref{DataLargeStrain}.}
\label{fig2}
\end{figure}

The linear rheological properties of the uncross-linked adhesives are characterized at $\omega=1$ rad/s (\textit{i.e.}\ at cyclic frequency $f=0.16$~Hz) in a parallel plate rheometer as a function of temperature (see Fig.\ \ref{fig2}). The measurements of the glass transition temperature $T_g$ are based on the inflection points of the $\ln(\mu')$ \textit{vs.}\ $T$ curves
and are reported in table \ref{Table1}. The change of the content of MA from 0 to 25\% results in the increase of $T_g$ by 20 \degree C.
The weak level of cross-linker, which is typical for PSA, does not affect the $T_g$ nor the linear rheology in the measured temperature range, but is expected to significantly change the maximum elongation that the adhesive can sustain before breaking or debonding, as it will be discussed in detail in section \ref{DataLargeStrain}.

All PSA tapes are peeled using three different experimental setups, to cover a broad velocity range, from 1 \textmu m/s to 4 m/s. These setups keep the peeling angle $\theta$ constant to a precision smaller than  2$\degree$ for steady-state peeling and smaller than 5$\degree$ when stick-slip occurs.
(1) For the lowest peeling velocities (1-100 \textmu m/s), PSA are simply deposited on a bar, which is then turned upside-down and inclined at a controlled angle. A weight  (setting $F$ very precisely) is attached to the free-standing part of the tape. The peeling velocity $V$ is precisely measured using time-lapse photographs of the peeling front.
(2) For the intermediate peeling velocities (from 10 \textmu m/s to 15~mm/s), we use an Instron testing machine (model 3343) to peel the PSA at a controlled velocity while recording the peeling force. In order to peel long  lengths of tape at a constant angle, the tape is deposited on an inclined bar translated (with a motor) at the same speed as the testing machine pulling velocity.
(3) Finally, the fastest peeling velocities (1 mm/s - 4 m/s) are imposed using a custom-made setup, where the tape is peeled at a constant angle from a horizontally translating bar, while being winded at the same velocity on a rotary motor equipped with a torquemeter to measure the applied force. The details of this setup can be found in a previous communication. \cite{Dalbe2014d}
The three setups are equipped with a lateral optical microscope allowing to image the structure of the debonding region with a micrometric resolution.

All adhesive tapes are  carefully bonded to the release side of a Scotch 600 tape and then peeled from it. When it is the Scotch 600 tape that is peeled, our protocol leads to experimental conditions similar to pre-existing measurements made by peeling directly from the roller. Another convenience of this protocol dwells in the moderate level of adhesion measured on the release side of Scotch 600 tapes: this results in peeling experiments with interfacial failure only, with no residuals on the substrate even at very low peeling velocities. In other words, no cohesive failure is observed in the experiments presented in this paper, which prevents an unnecessary complication of  the analysis due to cohesive-to-adhesive failure mode transition. \cite{Gent1969} Great care is taken with the release side of the Scotch 600 tape used as a substrate in order to avoid any damage of its release coating, which is critical to obtain reproducible results.

\section{Results}\label{Results}

\subsection{Dependence on the peeling angle}

\begin{figure}[tbp]
\centering
\includegraphics[width=8cm]{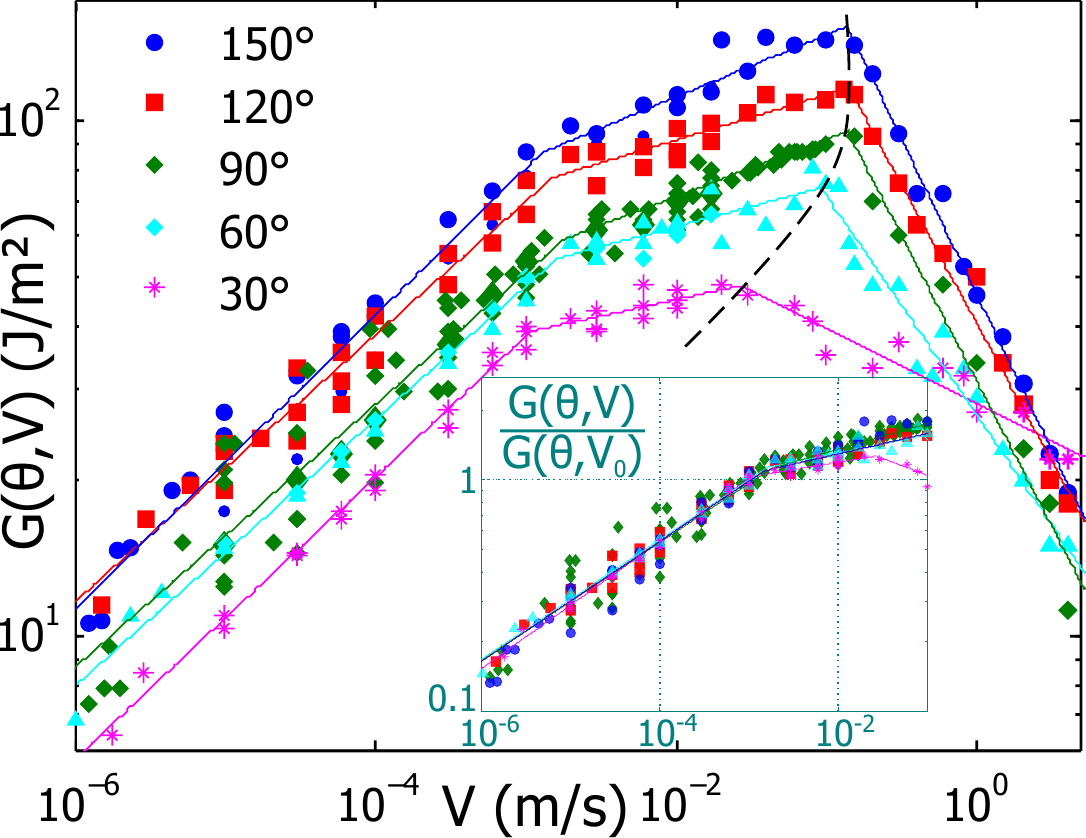}
 \caption{Measured energy release rates $G(V,\theta)$ for the 3M 600 Scotch tape for five different peeling angles. Straight lines represent power laws used as guides for the eye.
 The negative slope branches on the right of the dashed line correspond to stick-slip dynamics.
 Insert: master curve obtained by normalizing the $G(V)$ curve for each peeling angle by its average value at $V_0=1$~mm/s.
 }\label{fig3}
\end{figure}

According to the formalism of fracture mechanics and in the case of the peeling geometry, the energy release rate $G$ is directly related to the measured or imposed peeling force $F$:
\begin{equation}
G = \frac{F(1-\cos\theta)}{b}
\label{Rivlin},
\end{equation}
where $b$ is the tape width ($b=19$ mm for Scotch 600 and $b=12.5$ mm for all other studied tapes).
Expression~(\ref{Rivlin}) actually accounts for the work done by the force $F$ when the fracture grows by a unit surface, but discards the changes in the elastic energy stored in the tape backing, which are negligible for the considered tapes for peeling angles $\theta$ larger than 20\degree.\cite{Kendall1975} When peeling is steady, fracture mechanics assumes the balance between the energy release rate and the adherence energy, \textit{i.e.}\ $G=\Gamma$.
Fig.\ \ref{fig3} presents the energy release rate measurements for the peeling of Scotch 600 tapes at different  angles $\theta$ and velocities $V$.

For each peeling experiment, at least  several centimetres of tape are peeled, and up to several meters for the highest velocities. Each data couple $(V,G)$ reported in this article corresponds to a time average over the whole peeling experiment. One should note that, in the case of the highest peeling velocities (beyond the local maximum of $G(V)$), the peeling dynamics becomes unstable, resulting in large stick-slip velocity oscillations. This stick-slip dynamics makes the interpretation of the time average value of $G$ difficult, since $G$ cannot be viewed as a measurement of the adherence energy $\Gamma(V)$ over the unstable range of peeling velocities.\cite{Maugis88,Dalbe2014,Dalbe2014d}
Still, we leave these data on the plots (realised by averaging $G$ values over tens to hundreds of stick-slip cycles), in order to clearly locate the peak of dissipation and to show some remarkable features of the average $G(V)$ curves in the stick-slip domain, but they do not enter into our discussion about the mechanisms that determine  the adherence energy.

The consistency between the results obtained by the three experimental setups is assessed by checking that the $G(V)$ curves are superimposable in their overlapping velocity ranges. The validity of our experimental protocols is also assessed by the excellent agreement with the measurements of the $G(V)$ curve at $\theta=90\degree$ obtained in 1997 by Barquins and Ciccotti on a Scotch 600 tape (cf.\ Fig.\ 2 in Ref.\cite{Barquins1997}).

The $G(V)$ curves in this steady peeling domain appear similar for the different studied peeling angles $\theta$, with a global increase with  $\theta$: a clear dependence of $G$ (or $\Gamma$) on the peeling angle $\theta$ is demonstrated. The local maximum of the $G(V)$ curves is revealed to drift towards larger velocities for an increasing peeling angle $\theta$, in agreement with the results recently reported in Ref.\cite{Dalbe2014d} concerning the stick-slip threshold velocity.

The fact the distance in logarithmic scale between the $G(V,\theta)$ curves for different angles $\theta$ appears nearly constant as a function of the peeling velocity (in the steady peeling domain) reveals that $G(V,\theta)$ has separable dependences:
\begin{equation}
G(\theta,V)=f(\theta)g(V).
\label{Separability}
\end{equation}

This separability is even clearer in the insert of Fig.\ \ref{fig3}, where all $G(V,\theta)$ values are normalized by their value at $V_0=1$~mm/s (an arbitrary choice, in the middle of the examined velocity range).  This procedure reveals the velocity dependence $g(V)$  through a collapse of the data on a master curve  for velocities lower than $V=3$~cm/s, which is the onset of the stick-slip instability for $\theta=30\degree$.

 \begin{figure}[btp]
\centering
\includegraphics[height=6.2cm]{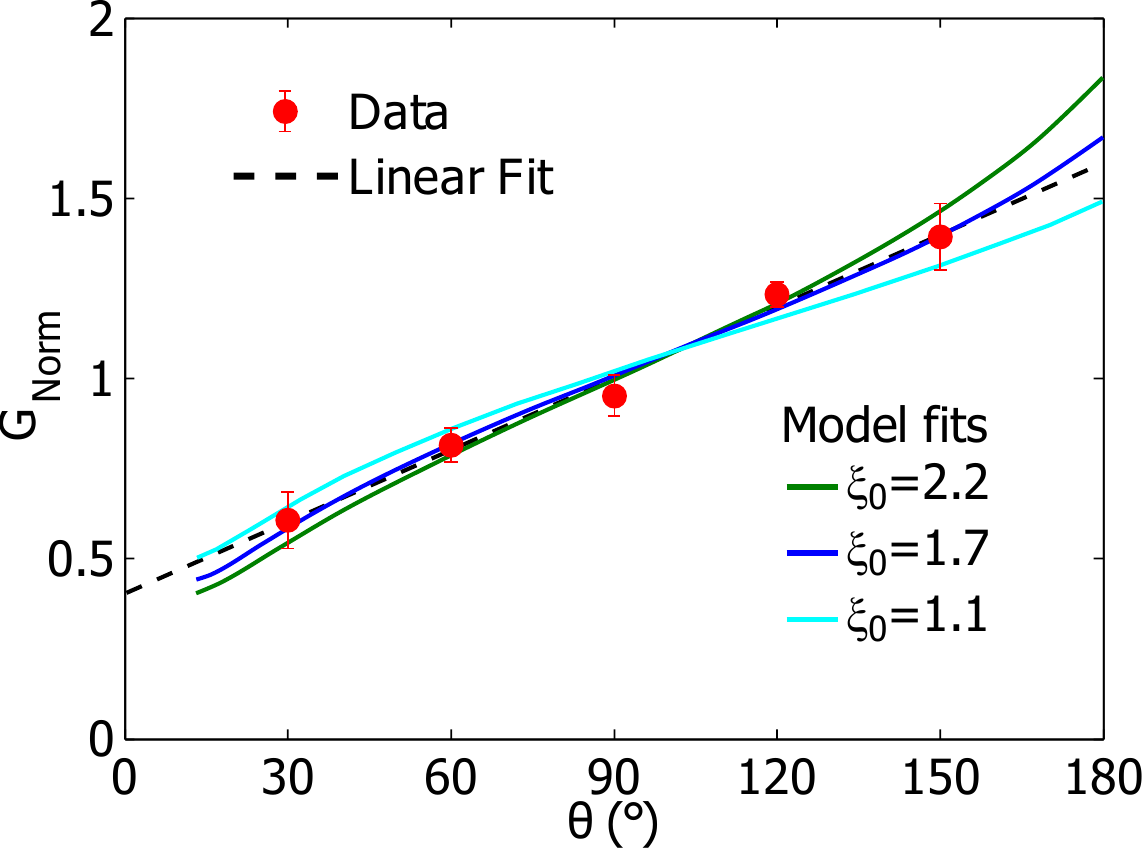}
 \caption{Master curve of the energy release rate dependence on the peeling angle, based on expression (\ref{Separability2}). Vertical error bars correspond to the standard deviations of $G/\left<G\right>_\theta$ when $V$ changes (see eq.\ (\ref{Separability2})). The three continuous lines represent theoretical fits according to eq. (\ref{G}) based on Kaelble's model (discussed in section \ref{Kaelble}).}
 \label{fig4}
\end{figure}

In order to isolate the angular dependence $f(\theta)$ of eq.\ (\ref{Separability}), we first evaluate the angular averaged velocity profile $\left<G\right>_\theta=\left<f\right>_\theta g(V)$,  which is expected to depend on the peeling velocity only, as follows :
\begin{equation}
\left<G\right>_\theta\approx \frac{1}{N_\theta}\sum_\theta G(\theta,V),
\end{equation}
where $N_\theta$ is  the number of regularly spaced studied peeling angles. We can eventually isolate $f(\theta)$ by computing $G/\left<G\right>_\theta$, which is equal to $f(\theta)/\left<f\right>_\theta$ according to eq.\ (\ref{Separability}). To improve the precision of our estimate of $f(\theta)$, we compute the average of $G/\langle G \rangle_\theta$ (which is indeed nearly independent on the peeling velocity) over the large number $N_V$ of studied peeling velocities:
\begin{equation}
\begin{split}
G_{Norm}(\theta)&=\left<\frac{G}{\left<G\right>_\theta}\right>_V\\
&\approx\frac{1}{N_V}\sum_V{\left(\frac{G(\theta,V)}{\left<G\right>_\theta}\right)}.
\end{split}
\label{Separability2}
\end{equation}
This estimate $G_{Norm}(\theta)$ is represented in Fig.\ \ref{fig4}. It increases with $\theta$ in an almost linear manner, catching the non-trivial dependence of the adherence energy $\Gamma$ with the peeling angle $\theta$.
The theoretical outcome of this dependence of $\Gamma$ on the loading geometry will be discussed in section \ref{Kaelble}.

\begin{figure}[btp]
\centering
\includegraphics[width=5.7cm]{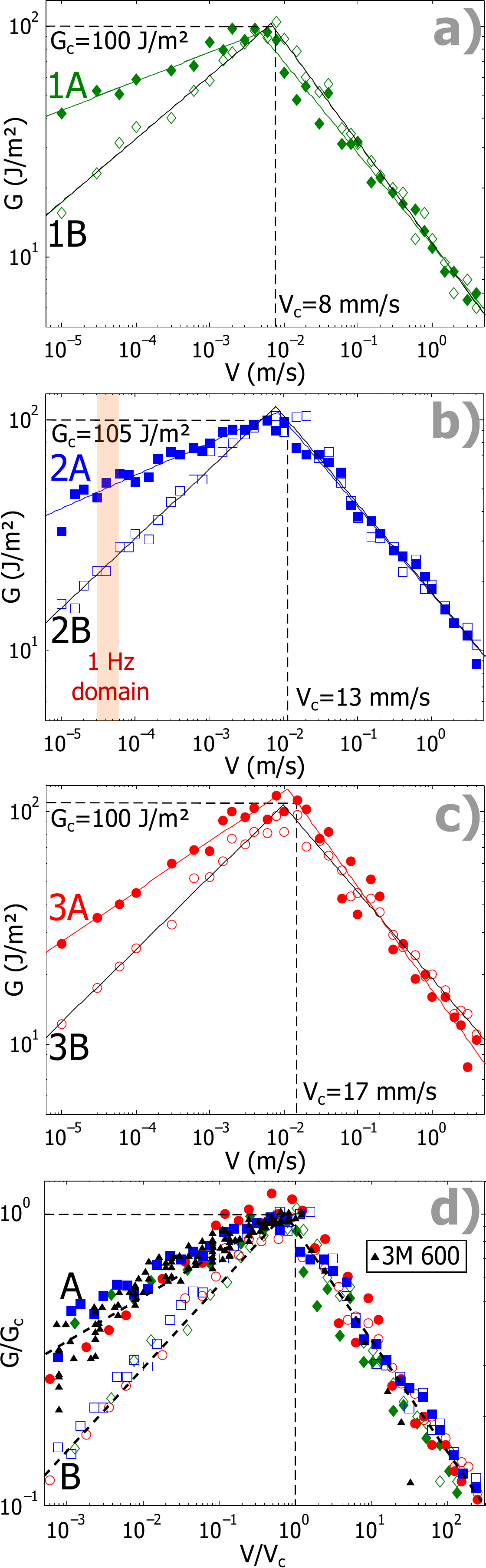}
\caption{(a)-(c): Measured energy release rates $G(V)$ at $\theta=90\degree$ for the six custom-made PSA presented in table \ref{Table1}. The pink region in (b) represents the peeling velocity associated to an effective strain rate of the adhesive around 1 Hz, as estimated from the microscopic observations of the debonding region in Fig.\ \ref{fig6}. In (d) master curves are represented, obtained by normalizing $G$ and $V$ by their values $G_c$ and $V_c$ at the onset of stick-slip, including the data of the Scotch 3M 600 at $\theta=90^\circ$.}
\label{fig5}
\end{figure}

\subsection{Dependence on the large strain rheology}
\label{DataLargeStrain}
In order to elucidate the different effects induced on the adherence energy by the changes of the linear and non linear rheology of the adhesive, we present
in Figs.\ \ref{fig5}(a), (b) and (c) the energy release rates $G(V)$ at room temperature $T=23\pm2\degree$C for the six custom-made adhesives presented in table \ref{Table1}.

They reveal that for each base composition, the increase of the level of cross-linker has the systematic effect of decreasing the adherence energy in the steady-state regime, while having no noticeable effect in the stick-slip domain. More precisely, in the steady-state regime,
the tapes with 0.4\% cross-linking level present a power-law $G\propto V^n$ with an exponent $n$ close to 0.3 ($n=0.29\pm0.02$ min/max),
while the tapes with 0.2\% cross-linking level follow a power-law with a significantly lower exponent ($n=0.175\pm0.045$ min/max).
In the stick-slip domain, all compositions present a power-law with an exponent very close to -0.4 ($n=-0.405\pm0.035$ min/max).

This systematic behaviour is even more striking in Fig.\ \ref{fig5} (d), where two distinct master curves are obtained when normalizing $G$ and $V$ by their values $G_c$ and $V_c$ at the onset of stick-slip, which coincide at first order with the local maximum of $G(V)$. To be more precise, there is a bistable domain close to the peak of the $G(V)$ curves, where the peeling alternates between steady-state and stick-slip dynamics, as reported in Ref.~\cite{Dalbe2014d} This domain affects less than a decade in peeling velocity. We estimate $(G_c,V_c)$ by the average of the $(G,V)$ measured data in this bistable domain, which is found to be quite a robust observable.
Fig.\ \ref{fig5} (d) shows that the behaviour of the less cross-linked custom-made adhesives tested is close to that of Scotch 600; moreover, data from the literature \cite{Maugis88} show that another commercial tape, Scotch 3M 602, follows an increasing power-law with an exponent $n=0.35$, which is just a little higher than the one of our more cross-linked adhesives: all our PSA are therefore truly representative of commercial PSA used as office tape.

For the six custom-made adhesives, no change in $G_c$ is observed beyond data scattering, meaning that the vertical shifts $1/G_c$ used in Fig.\ \ref{fig5} (d) are essentially the same.
The horizontal shift factors correspond to a change of $V_c$ of about a factor 2 when increasing the content of MA from 0 to 25\%. These shift factors are expected to be a consequence of the temperature shifts of the rheological properties with composition, illustrated in Fig.\ \ref{fig2}, combined with the
acknowledged time-temperature equivalence of the rheology of polymers,\cite{Ferry1970} which is known to be reflected on the adherence curves of PSA (see for example Fig.\ 9 in  \cite{Derail1998} or more generally Refs. \cite{Kaelble1964,Gent1969,Maugis1985,Derail1998}).

In order to appropriately apply the time-temperature equivalence, we should first identify the frequency range solicited by the peeling at ambient temperature over the three decades of measured steady-state peeling velocities $V$: $10^{-5}$ to $10^{-2}$ m/s. According to the Cox-Merz rule,\cite{Cox1958} we estimate the relevant frequencies $f^*$ by the strain rates $\dot\varepsilon$ experienced by the adhesive in the debonding region, namely:

\begin{equation}
f^*\sim\dot\varepsilon\sim \varepsilon_{max} V/L_{dr}
\label{CoxMerz},
\end{equation}
where $L_{dr}$ and $\varepsilon_{max}$ are respectively the size of the debonding region (see Fig.\ \ref{fig1}) and the maximum strain experienced by the adhesive fibrils obtained through the microscopic observations presented in section \ref{Microanalysis}. We find that they are between 0.2 and 100 Hz at 23\degree C (the 1 Hz domain is plotted on Fig.\ \ref{fig5} (b) for reference).
In order to transpose this frequency range on the rheological measurements of Fig.\ \ref{fig2}, we can use the time-temperature superposition principle. To a first rough approximation, increasing the strain frequency by one decade corresponds to decreasing the temperature by 5 to 10\degree C. We thus find that the relevant frequency range at 23\degree C corresponds to a temperature range at 1 rad/s (0.16 Hz) of about 15-30\degree C below room temperature. The solicited part of the linear rheology is thus well included in the entanglement plateau, which corresponds to the low moduli zone of Fig.\ \ref{fig2} below 1 MPa. We remark that the weak level of cross-linking of PSA does not affect the linear rheology in this part of the entanglement plateau, but only at much higher temperatures, or equivalently at much lower peeling velocities.

Since the relevant temperature range is higher than the glass transition temperature of our custom-made adhesives by more than $50\degree$C,
the temperature shifts $\Delta T=T_j-T_i$ estimated in Fig.\ \ref{fig2} (b) and (d) for the adhesives $i$ and $j$ should be related to the corresponding peeling velocity shifts $1/V_{ci}$ and $1/V_{cj}$ estimated in Fig.\ \ref{fig5} by an Arrhenius-like law:\cite{Ferry1970}
\begin{equation}
\ln\left(\frac{a_{Tj}}{a_{Ti}}\right)=
\ln\left(\frac{V_{ci}}{V_{cj}}\right)=\frac{E_a}{R}\left(\frac{1}{T_j}-\frac{1}{T_i}\right),
\label{Arrhenius2}
\end{equation}
where $E_a$ represents a typical activation energy and $R$ is the universal gas constant.

Although these temperature and velocity shifts are small, we find a good correlation between these two types of shifts when testing eq.\ (\ref{Arrhenius2}), with a linear correlation coefficient of $0.988$. The measured activation energy $E_a$ is found to be between 40 and 50 kJ/mol for our adhesives.
This activation energy  is a little lower than what has been recently measured on pure poly (n-butyl acrylate), a  polymer typically used for PSA, namely $60$ to $80$ kJ/mol. \cite{Callies2015}

The horizontal reascaling of the peeling velocity by $1/V_c$ can thus clearly be attributed to the temperature shifts in linear rheology induced by the change in the content of MA.
Nevertheless, the collapse of all measured peel curves on two different master curves, depending only on the level of cross-linking, clearly demonstrates that the sole linear rheology is insufficient to describe or predict the adherence energy $\Gamma(V)$ over the whole range of  peeling velocities. Notice however that the influence of cross-linking on the adherence energy is progressively reduced when increasing the peeling velocity and is no longer detectable close and beyond the $\Gamma(V)$ maximum.

\subsection{Microscopic investigation of the debonding region}
\label{Microanalysis}

\begin{figure}[t]
\centering
\includegraphics[width=8cm]{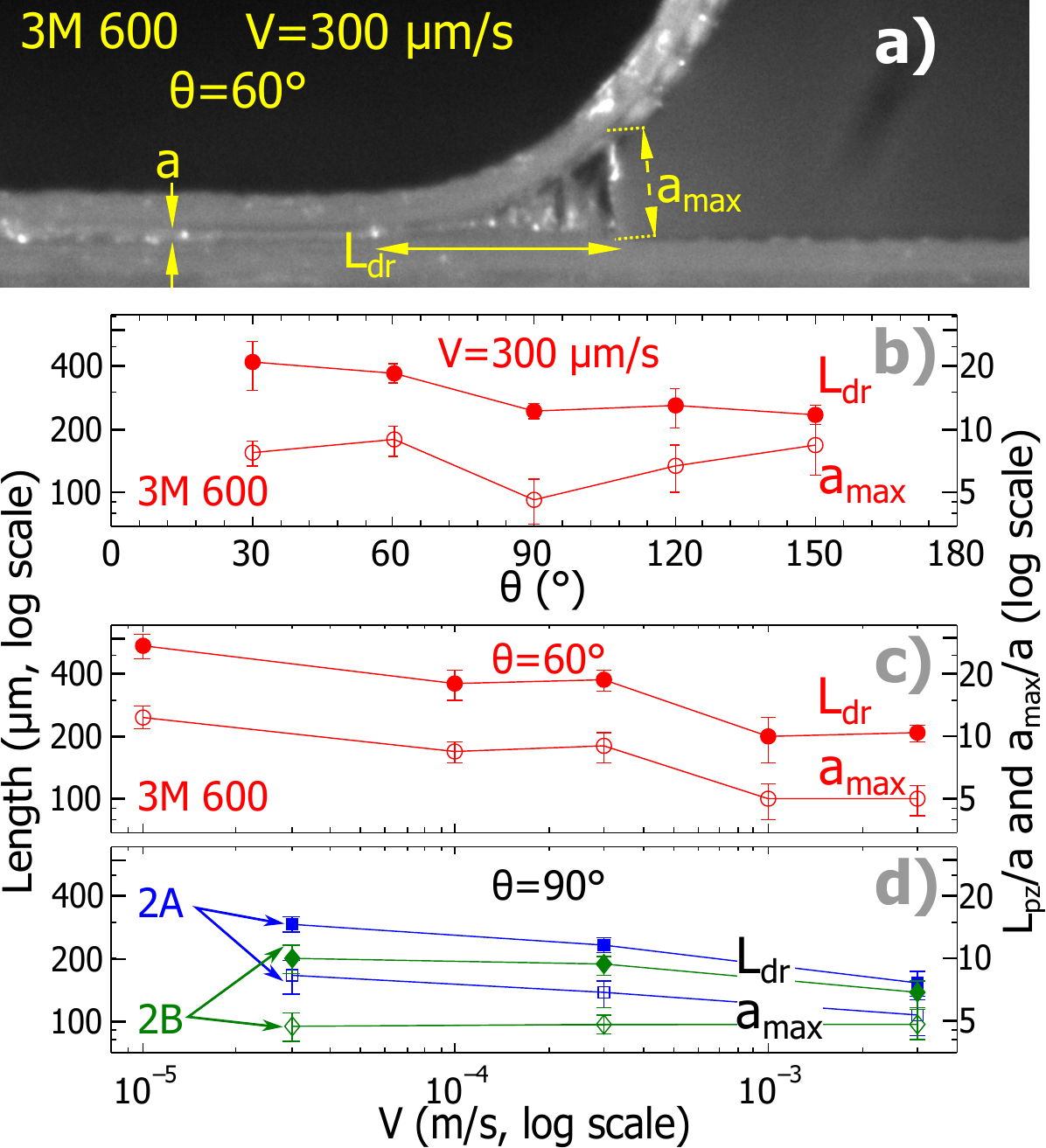}
 \caption{Geometrical parameters $a_{max}$ (empty symbols) and $L_{dr}$ (full symbols) of the debonding region. For the three charts, the right ordinate is equal to the left one divided by the adhesive thickness ($a=20$ \textmu m), to clearly represent the stretch ratio $a_{max}/a$. Abscissa and ordinate are in log scale to emphasize the almost constant ratio $a_{max}/L_{dr}$ when $V$ is changed. }
 \label{fig6}
\end{figure}

Fig.\ \ref{fig6} (a) presents a typical image of the debonding region during steady-state peeling where the characteristic size $L_{dr}$ and the length $a_{max}$ of the longest glue fibril at the peeling front are extracted by image analysis and averaged over the length of the movie (see for example the one given in the supplementary materials of this paper).

Fig.\ \ref{fig6} (b) does not reveal a clear dependence of the fibrils length $a_{max}$ on the peeling angle for the Scotch 3M 600. On the contrary, the length of the debonding region $L_{dr}$ is shown to slowly decrease with increasing $\theta$. This is qualitatively consistent with the measured simultaneous increase of the peeling force $F$: since the tape backing radius of curvature $R_c$ at the peeling front scales with $F^{-1/2}$ ($R_c\sim\sqrt{EI/F(1-\cos\theta)}$ with $EI$ the bending modulus of the tape backing), the tape should be slightly more curved as the peeling angle increases. If the maximum stretch of the fibrils is independent of this curvature, then for geometric reasons the length $L_{dr}$ of the region where the adhesive is significantly stretched should be slightly shorter when $\theta$ increases.

Fig.\ \ref{fig6} (c) and (d) show that $a_{max}$ and $L_{dr}$ globally decrease with increasing peeling velocities $V$, except for the length $a_{max}$ of the last fibrils for the most cross-linked adhesive 2B, which seems to be insensitive to $V$. Moreover, the ratio $a_{max}/L_{dr}$ is almost constant with $V$, which means that the geometry of the zone where the adhesive is stretched is only downscaled when $V$ is increased, at least in the steady peeling regime.

Fig.\ \ref{fig6} (d) provides a very interesting insight on the role of cross-linking: it shows that, at low enough peeling velocities, the size $L_{dr}$ of the debonding region, and especially the length  $a_{max}$  of the longest glue fibril at the peeling front, are considerably smaller for the more cross-linked adhesives, even if these adhesives have the same linear rheology. Since an adhesive with a shorter chain length between cross-links is less able to be stretched before debonding or breaking, this observation actually seems  logical, but its impact on the adherence energy was not investigated before and was not clearly decoupled from other properties such as linear rheology.

Moreover, we observe a progressive collapse for the dimensions of the fibrillated zone when the peeling velocity $V$ approaches the local maximum of $\Gamma$, which provides a sound rationale for the observation that the effect of the cross-linking on $\Gamma$ is progressively lost close and beyond the $\Gamma(V)$ maximum, as it will be further discussed in section \ref{LargeStrains}. More precisely, the ratio between the $\Gamma$ values for the two cross-linking levels
is actually well correlated (linear correlation coefficient $0.9987$) to the ratio between the lengths $a_{max}$. For example, the ratios $\Gamma(2A)/\Gamma(2B)$ are 2.25 at 30 \textmu m/s, 1.55 at 300 \textmu m/s and 1.05 at 3 mm/s  (see Fig.\ \ref{fig5}), while the ratios $a_{max}(2A)/a_{max}(2B)$ are 1.75, 1.4 and 1.1 at the same respective peeling velocities.

\section{Interpretation and discussion}\label{Discussion}
\subsection{Choice of the interpretation model family}
For  an unconfined soft elastic material, the radius of curvature of the stressed crack tip can become much larger than inter-molecular distances, namely in the \textmu m to mm range for the materials  used in PSA.
This radius  is given in order of magnitude by the elasto-adhesive length $\ell_{ea}=\Gamma/Y$ (where $Y$ is the Young's modulus of the adhesive).  Hui \textit{et al.}\cite{Hui2003} demonstrated that the LEFM stress singularity is cut off at this distance $\ell_{ea}$ from the crack tip, causing a saturation of stress (at $2Y$ in the case of a neo-Hookean material): this is the so-called ``elastic blunting" phenomenon. During the steady-state peeling of a PSA, $\Gamma$ increases typically from 10 to 100 J/m$^2$  with the peeling velocity (\textit{cf.} Figs. \ref{fig3} and \ref{fig5}) and the storage modulus concomitantly increases from tens to hundreds of kPa, in the relevant entanglement plateau domain (\textit{cf.} Fig.\ \ref{fig2}),   because of the increasing  strain rate $\dot\varepsilon \sim V\varepsilon_{max}/L_{dr}$ and  of viscoelasticity. Therefore, the order of magnitude of $\ell_{ea}$ in the steady-state peeling branch is of several hundreds of \textmu m, which is  larger than the typical thickness of the adhesive layer $a\sim20$\textmu m.

A second peculiarity of soft dense materials is their substantial incompressibility (the bulk modulus $K$ is typically four to five orders of magnitude larger than the shear modulus $\mu$). Large volume expansions are thus not possible without developing cavitation. The stress criteria for cavitation have been extensively investigated:\cite{Gent1959,Dollhofer2004,Chiche2005} they mainly depend on hydrostatic negative pressure, but also on stress triaxiality. In the particular case of an isotropic tension, the cavitation stress threshold is minimum and  close to the Young's modulus $Y$.\cite{Gent1972} Even under uniaxial traction of a thin confined film, incompressibility implies the development of hydrostatic tension to which the film responds in an \oe dometric way. When this kind of loading geometry  is applied, the relevant modulus  would approach the bulk modulus far from the film lateral boundaries\footnote[3]{More precisely, the relevant modulus would be the so-called \oe dometric or longitudinal wave  bulk modulus $\tilde Y=[Y(1-\nu)]/[(1-2\nu)(1+\nu)]$  with $Y$ the Young's modulus of the adhesive and $\nu$ its Poisson's ratio, see Ref. \cite{Landau1986}}. Experimentally, the response is critically dependent on the deviation from perfect incompressibility as well as on material defects, which lead to an effective modulus one or two orders of magnitude above $Y$.\cite{Chiche2003,Chiche2005}

When dealing with a strongly confined soft  material, as in the peeling of  PSA, these  peculiarities (thickness $a$ small compared to other lateral dimensions and to $\ell_{ea}$ and incompressibility) lead to a dramatic change of the fracture mechanisms compared to an unconfined and/or hard material. The first consequence comes from the geometric confinement (lateral dimensions large compared to the adhesive thickness $a$) and from the Saint-Venant principle: far from the peeling front, stress can be considered as uniform through the bond thickness $a$, from the tape backing to the substrate. Moreover, its lateral variations  are correlated over a distance of the order of $a$. In this region, the adhesive can thus be treated as if it were divided into strands separated (in both directions perpendicular to the tape thickness) by a distance of the order of  $a$.

The second consequence comes from the so-called ``elasto-adhesive confinement": the fact that $a\ll\ell_{ea}$ prevents the development of the LEFM stress singularity inside the adhesive. The stress distribution tends therefore to be constant through the thickness of the bond, even close to the peeling front.\cite{Hui2003} Moreover, it has been acknowledged since the earliest studies that the large adhesion of PSA is related to the occurrence of large extensions of the adhesive before debonding, which has been shown to necessarily occur through cavitation and stringing: in this region close to the peeling front, long fibrils are present.

These effects of confinement, of incompressibility and of stringing lead to a description of the debonding region divided in two domains (\textit{cf.}\ Fig.\ \ref{fig1}): an inner one before cavitation where the response of the adhesive is stiffer than the Young's modulus, due to incompressibility, and an outer fibrillated domain where the response is essentially uniaxial and unconfined, thus comparable to the Young's modulus. The theoretical arguments and experimental observations (the presence of fibrils) mentioned above justify that both domains can  be considered as arrays of parallel strands experiencing extension (which is natural for the fibrillated part), providing an effective cohesive zone behaviour. The respective contributions of each of these two domains to the global mechanical response measured in a peeling experiment may however not be easy to untangle. Therefore, the link between this global response and the material properties is a challenging issue.

These conceptual arguments provide a sound justification for the use of the foundation-based   family of models, which describe the debonding region  as a parallel array of strands, represented by (possibly non-linear) springs and dashpots,  coupling the flexible tape backing to the substrate. In this approach, viscous friction occurs within the dashpots, but the most important source of  energy dissipation is elastic hysteresis: the work used to deform a spring is entirely lost when it either breaks or debonds. On the contrary, the other type of modelling  presented in the introduction  is clearly not relevant to confined soft adhesives, since it is based on a viscoelastic perturbation of the LEFM singularity which is not developed at all within the scale of the adhesive thickness. We therefore focus on the foundation-based  category of models. Even if they use a simplified description of the strands response, they can catch some of the key mechanisms that set the stress and strain distributions within the adhesive, which in turn should set the adherence energy. Some models among this category  allow in particular to account for the possible influence of the large strain rheology of the adhesive material on the adherence energy. \cite{Gent1969}

\subsection{Bond stress distribution and angular dependence}\label{Kaelble}

Kaelble's mechanical description of the debonding region during peeling \cite{Kaelble1960} is the first (chronologically) and is certainly a reference model within the foundation-based approach. His linear description of the adhesive viscoelastic response eventually allows for a full analytical solution of the bond stress distribution, with  one adjustable parameter only: the critical stress at debonding. A  direct measurement of this critical stress is however  a challenging experimental issue.

The predictive nature of this model allows quantitative comparisons with experiments: it predicts in particular a non-trivial dependence  of the adherence energy $\Gamma$ on the adhesive thickness $a$ \cite{Kaelble1992} and on the peeling angle $\theta$. The latter prediction should be tested, since this dependence, while indirectly present in already published data,\cite{Kaelble1960}  has been never addressed in detail nor  physically interpreted, even by Kaelble himself.

\begin{figure}[tbp]
\centering
\includegraphics[width=6cm]{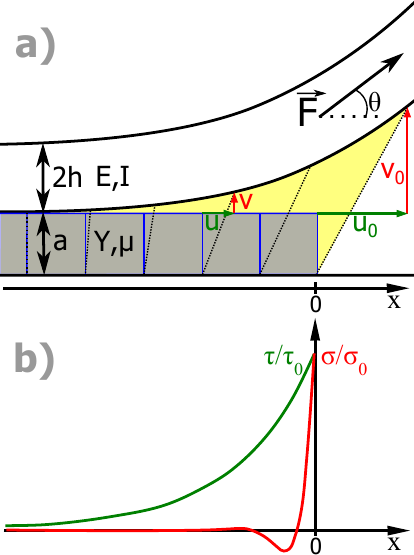}
 \caption{(a): Diagram of Kaelble's modelling of the debonding region. The adhesive, initially unstrained (in gray) is sheared and stretched by the peeling force, transmitted through the stretch and bending of the tape backing. (b): Typical shear (green) and stretch (red) stresses in the adhesive as a function of the longitudinal distance $x$ to the peeling front,
according to eq.\ (\ref{Stress}).}
\label{fig7}
\end{figure}

Let us summarize the assumptions and main predictions of this model. The key ingredient is the transmission of the peeling force $F$ to the adhesive through the tape backing elongation  and bending, leading respectively to  shear $\tau(x)$ and cleavage (or stretch) $\sigma(x)$ stress distributions inside the adhesive, homogeneous through the adhesive width and thickness and only dependent on the  longitudinal coordinate  $x$  along the  debonding region (see Fig.\ \ref{fig7}).  As in every model within the foundation-based approach, the adhesive is modelled by individual strands that link the tape backing to the (rigid) substrate and that can be sheared and stretched independently up to debonding. In Kaelble's model, these elements are linear and elastic with a Young's modulus $Y$ and a shear modulus $\mu$. To a first approximation, viscoelasticity can   be  accounted for   by considering the increase of $Y$ and $\mu$ with strain rate, which is controlled by the peeling velocity: this model thus rather considers rate-dependent elasticity than true viscoelasticity, viscous dissipation being neglected.

The physics of this model can be easily understood using simple scaling laws: shear in the adhesive is determined by the progressive transfer of the stretch energy of the backing  (of Young's modulus $E$ and thickness $2h$) to the adhesive,  along the characteristic length $\lambda_\alpha$. This length  can therefore be determined by comparing the  longitudinal stretch energy $U_{Str}^{B}$ of the tape backing with the shear energy $U_{Sh}^{A}$ of the adhesive. If these energies are associated to a characteristic horizontal displacement $u_0$ (see Fig.\ \ref{fig7}), we obtain:
\begin{equation}
 U_{Str}^{B}\sim\left(E\frac{u_0^2}{\lambda_\alpha^2}\right)hb\lambda_\alpha\sim U_{Sh}^{A}\sim\left(\mu\frac{u_0^2}{a^2}\right)ab\lambda_\alpha .
\label{ScalingAlpha1}
\end{equation}
When considering typical  geometrical and mechanical characteristics of   PSA, \textit{i.e.} $a\sim h$ and $E/\mu\sim10^4$, this scaling analysis leads to:
\begin{equation}
\lambda_\alpha\sim\sqrt{\frac{Eah}{\mu} }\sim100a.
\label{ScalingAlpha2}
\end{equation}

Similarly, cleavage in the adhesive is determined by the progressive transfer of the bending energy of the backing (of bending modulus $EI$, where $I=2bh^3/3$) to the adhesive,  along the characteristic length $\lambda_\beta$. This length  can therefore be determined by comparing the bending energy $U_{Bend}^{B}$ of the tape backing with the stretch energy $U_{Str}^{A}$ of the adhesive. If these energies are associated to a characteristic vertical  displacement $v_0$ (see Fig.\ \ref{fig7}), we obtain:
\begin{equation}
\begin{split}
U_{Bend}^{B}\sim&\left(\frac{EI}{R_c^2}\right)\lambda_\beta \sim \left(\frac{EIv_0^2}{\lambda_\beta^4}\right)\lambda_\beta\\
&\sim U_{Str}^{A}\sim\left(Y\frac{v_0^2}{a^2}\right)ab\lambda_\beta ,
\end{split}
\label{ScalingBeta1}
\end{equation}
where $R_c\sim v_0/\lambda_\beta^2$ is the typical radius of curvature of the tape backing. Using the same typical   characteristics of PSA as in the evaluation of $\lambda_\alpha$, this scaling analysis leads to:
\begin{equation}
\lambda_\beta\sim\sqrt[4]{\frac{EIa}{Yb}}\sim10a.
\label{ScalingBeta2}
\end{equation}

One can notice that stretch of the adhesive is much more concentrated than shear, due to the two different characteristic scales of stress concentration $\lambda_\alpha$ and $\lambda_\beta$. These  lengths are independent from the loading conditions, which on the contrary set the typical stresses $\sigma_0$ and $\tau_0$ in the adhesive close to the peeling front:
\begin{equation}
F\cos\theta\sim b\lambda_\alpha\tau_0 \qquad ; \qquad F\sin\theta\sim b\lambda_\beta\sigma_0.
\label{ScalingF}
\end{equation}

To go beyond these scaling laws, one needs to write the complete equations of static equilibrium of forces and moments (\textit{cf.} Ref. \cite{Kaelble1960}), which lead to an exact analytical solution for the two stress distributions $\tau(x)$ and $\sigma(x)$ (see Fig.\ \ref{fig7} (b)):
\begin{equation}
\tau=\tau_0 e^{\alpha x} \qquad \qquad  \sigma=\sigma_0 e^{\beta x}\left(\cos\beta x +K sin\beta x\right),
\label{Stress}
\end{equation}
\begin{center}
with
\end{center}
\begin{equation}
\begin{split}
&\tau_0=\frac{\alpha}{b}F\cos\theta \qquad \: \sigma_0=\frac{2\beta}{b\left(1-K(F,\theta)\right)}F\sin\theta\\
&\alpha=\sqrt{\frac{\mu}{2Eah}} \qquad\ \: \beta=\sqrt[4]{\frac{Yb}{4EIa}}=\sqrt[4]{\frac{3Y}{8Eah^3}}\\
&K=1-\frac{\sin\theta}{\beta\sqrt{2EI(1-\cos\theta)/F}-h\beta\cos\theta+\sin\theta}.\\
\end{split}
\label{Stress2}
\end{equation}
where $\tau_0$ and $\sigma_0$ represent the maximum shear and cleavage stresses at the peeling front (the typical stresses in the  scaling-law approximation of eq.\ (\ref{ScalingF})).
For 3M Scotch 600, $2h=38$ \textmu m, $E=1.26$ GPa, $I=87\cdot 10^{-18}$ m$^4$, $a=20$ \textmu m, $b=19$~mm and $Y=3\mu$ is in the tens to hundreds of kPa range, depending on the peeling velocity.

As predicted from scaling laws (\ref{ScalingAlpha2}) and (\ref{ScalingBeta2}), the two stress distributions are concentrated over  different characteristic lengths, $\lambda_\alpha=1/\alpha$ and $\lambda_\beta=1/\beta$ for shear and cleavage respectively. While  shear stress follows a simple exponential decay, cleavage stress  decay is actually modulated by an oscillation, both having the same  characteristic length $\lambda_\beta$, which has been confirmed experimentally.\cite{Kaelble1960,Niesiolowski1981} The dimensionless parameter $K$ describes the phase shift of this oscillation, which is set by the ratio between shear and cleavage loadings, \textit{i.e.} by the peeling force and angle.
While the three scaling laws (\ref{ScalingAlpha2}), (\ref{ScalingBeta2}) and (\ref{ScalingF}) capture the main dependences of $\tau_0$ and $\sigma_0$ written in eq.\ (\ref{Stress2}), they lose the detail of the phase shift in the cleavage stress distribution represented by the dimensionless parameter $K$.

The set of expressions (\ref{Stress}) and (\ref{Stress2}) only describes a static equilibrium: we therefore need a criterion for  the peeling front to move. Such a criterion can be met  when either the shear or cleavage stress reach a critical value, $\tau_c$ or $\sigma_c$ respectively. In this case, the peeling force $F$ can be calculated from eq.\ (\ref{Stress2})  by setting $\tau_0$ or $\sigma_0$ to this critical value. The adherence energy $\Gamma$ can finally be calculated through eq.\ (\ref{Rivlin}), since $\Gamma=G$ when the peeling is steady.
For our  experimental parameters, namely since the peeling angle is not too close to $0\degree$, $\sigma_0$ is predicted to be at least two orders of magnitude higher than $\tau_0$, even around $180\degree$. We may therefore reasonably think that the cleavage stress criterion is the first to be fulfilled and that the adherence energy calculation should be based on $\sigma_0=\sigma_c$. Incidentally, if the critical shear criterion were reached, this would rather induce  sliding than debonding, as shown by Chaudhury \textit{et al.},\cite{Newby1997} which is never observed in our microscopic films (see for example the one given in the supplementary materials of this paper).

Due to the implicit form of eq.\ (\ref{Stress2}), Kaelble proposed a numerical solution, but we were able to derive an exact analytical solution\footnote[2]{This solution is  obtained by extracting $F$ from the $\sigma_0(=\sigma_c)$ expression in eq.\  (\ref{Stress2}) and by replacing $K$ with its explicit expression. The resulting implicit equation can then be rearranged into a simple biquadratic  equation which possesses  one   positive real solution only.}
which allows for appreciating the explicit dependence of the fracture energy $\Gamma$ on the peeling angle $\theta$:
 \begin{equation}
\Gamma=a\mathcal{W}K'^2(\xi)
\label{G}
\end{equation}
\begin{center}
with
\end{center}
 \begin{equation}
 \begin{split}
&\mathcal{W}=\frac{\sigma_c^2}{2Y}\qquad ; \qquad  K'(\xi)=\frac{2}{\xi}\left(1-\sqrt{1+\xi}\right)\\
&\xi=\xi_0\left(\frac{\sin\theta-h\beta\cos\theta}{1-\cos\theta}\right ) \quad ; \quad \xi_0=4a\frac{\sigma_c}{Y}\beta.
\end{split}
\label{G2}
\end{equation}
We introduce the $K'$ notation to make a distinction with Kaelble's $K$ parameter, even if $K'$ and $K$ are actually very close in our experimental parameters range.

At first order, the predicted adherence energy is proportional to the adhesive thickness $a$\footnote[4]{$K'^2$ also depends on $a$, but very weakly: when considering typical geometrical and mechanical characteristics of PSA, $K'^2$ only decreases by a factor 2 when $a$ increases from 1 to 100 \textmu m.
} and to the volume density of stored elastic energy  $\mathcal{W}$ just before debonding of each individual strand. The angular dependence embedded in the dimensionless parameter $K'$ can be physically  understood by the following argument: an increase of the peeling angle has the effect of shifting the cleavage stress oscillation (see Fig.\ \ref{fig7} (b)) closer to the peeling front. This increases the contribution of the compressive forces into both the resultant force and moment. If the cleavage critical stress $\sigma_c$ is held constant, an increase of the peeling angle results in a decrease of the peeling force which does not simply correspond to the $1/(1-\cos\theta)$ geometric term implied by  eq.\ (\ref{Rivlin}). It  induces a net increase in the apparent fracture energy $\Gamma$ with the peeling angle.

This increase is indeed experimentally observed in our data (see Fig.\ \ref{fig3}). The observed separability between peeling velocity and angular dependences enables us to focus on the latter, as it is done in Fig.\ \ref{fig4}. But this separability, experimentally observed, needs to be justified from a theoretical point of view. Equation (\ref{G}) does predict   a partial separability between the peeling velocity dependence, mainly due to  $\mathcal{W}(V)$, and $\xi$, which in turn depends on $\theta$. However, $\xi$  also depends on the two dimensionless parameters $\xi_0=4a\beta\sigma_c/Y$ and $h\beta$, both depending on $V$ through $\sigma_c$ and $Y$. This can be summarized in:
\begin{equation}
\Gamma=a\mathcal{W}(V)\times K'^2\left(\theta,\xi_0(V),h\beta(V)\right).
\label{SeparabilityTheo1}
\end{equation}
For the peeling velocities tested, $Y$ is expected to change by a factor less than 10 (from tens to hundreds of kPa, in the relevant entanglement plateau),\cite{Lakrout1999,Lindner2007} so $\beta\propto Y^{1/4}$ can be assumed to be nearly constant. Moreover, if $\xi_0=4a\beta\sigma_c/Y$ changes, it should  be  because  of $\sigma_c/Y$. The only sensible dependences of $K'^2$ are therefore  on $\theta$ and   $\sigma_c/Y$:
\begin{equation}
\begin{split}
\Gamma&\approx a\mathcal{W}(V)\times K'^2\left(\theta,\frac{\sigma_c(V)}{Y(V)}\right)\\
&\approx g(V)\times f\left(\theta,\frac{\sigma_c(V)}{Y(V)}\right).
\end{split}
\label{SeparabilityTheo2}
\end{equation}
This expression can be compared to eq.\ (\ref{Separability}):
the experimental separability between the peeling angle and velocity means, if Kaelble's interpretation of the adherence energy is correct, that $\sigma_c/Y$ is almost independent of the peeling velocity for the studied adhesive tape Scotch 600  in the range $V=1-10^4$~\textmu m/s.

The physical meaning of a change of $\sigma_c$ with the peeling velocity is still unclear. However, the estimates reported by Kaelble (based on measurements of Ref.~\cite{Smith1958}) are consistent with a constant ratio $\sigma_c/Y$ over the 4 decades of steady-state peeling velocity below the onset of stick-slip (\textit{cf.} Fig.\ 1 in Ref.~\cite{Kaelble1964}).

Finally, we can quantitatively compare the angular dependence predicted by eq.\ (\ref{G}) to our data, as it is done in Fig.\ \ref{fig4}. To do so, we compute the theoretical value of the adherence energy $\Gamma_{th}$ from eq.\ (\ref{G}) for each peeling angle and for different values of $Y$ and $\sigma_c/Y$. We then compute $G_{Norm}=f(\theta)/\left<f\right>_\theta$ as defined by eq.\ (\ref{Separability2}), which has the main function of suppressing the $g(V)$ (or $a\mathcal{W}(V)$) pre-factor:
\begin{equation}
G_{Norm,th}=\frac{\Gamma_{th}}{\left<\Gamma_{th}\right>_\theta},
\end{equation}
where the angular average is performed over the five peeling angles experimentally studied, as it is also done  for   the experimental data. We confirm that this estimate $G_{Norm,th}$ is essentially independent of $Y$ (as long as $Y$ stays in the entanglement plateau values, in the tens to hundreds of kPa range), as predicted in the second-to-last precedent paragraph (containing eqs. (\ref{SeparabilityTheo1}) and (\ref{SeparabilityTheo2})). The only detectable dependences of $G_{Norm,th}$ are indeed on $\theta$ and on $\sigma_c/Y$, that is to say on $\theta$ and on $\xi_0$. Fig.\ \ref{fig4} shows that the angular dependence of the adherence energy is well reproduced
when choosing $\xi_0=1.7 \pm 0.5$.

This value $\xi_0=1.7$  corresponds to $\sigma_c/Y\sim3.5-4$, assuming $Y$ in the tens to hundreds of kPa range.  $\sigma_c$ would thus be  in the hundreds of kPa to MPa range, which is indeed the typical maximum pressure observed in probe-tack experiments on adhesives similar to those used in PSA, and which is also consistent with  the stresses measured during peeling experiments.\cite{Kaelble1960,Kaelble1964,Kaelble1965,Niesiolowski1981} Notice that Kaelble's data also imply a similar value of $\sigma_c/Y$, which reaches 4.5 (as can be extracted from Fig.\ 1 (B) in Ref.~\cite{Kaelble1964}).

However, the straightforward interpretation of $\xi_0$ leads to an apparent contradiction: since Kaelble's model is intrinsically linear, the ratio $\sigma_c/Y$ should be interpreted as a maximum deformation,  around 350-400\% ! This is way too large for a linear response. While Kaelble acknowledged the occurrence of fibrillation, he neglected the influence of the large stretched region of the adhesive in his stress equilibrium analysis, because it would have prevented his analytical treatment. One should therefore not try to give a completely quantitative interpretation of $\xi_0$ in terms of bulk linear parameters of the adhesive. However, we   emphasize the robustness of  Kaelble's model to describe the peeling angle dependence of the adherence energy, even if, in a fully realistic model, $\Gamma$  should  be related to the bond stress distribution in a more general way: the complete behaviour of the adhesive, beyond the linear response, should be taken into account.

Although Kaelble's model was conceived to predict the viscoelastic energy dissipation associated to peeling, it leads to dissipation even in a purely  elastic case, which is in apparent contradiction with Griffith's energy balance. This paradox can be solved by noting that the representation of a confined and soft adhesive by an elastic foundation (\textit{i.e.} a parallel array of independent springs) does not correspond to an elastic continuum, such as in Griffith's theory. The independent failure of the strands indeed leads to an energy loss by  elastic hysteresis (dependent of the peeling rate because the Young's modulus $Y$ changes    with $V$), even if the strands are purely Hookean. This description is actually even more relevant for the fibrillated part of the adhesive, where the independent failure of the fibrils is apparent in our microscopic imaging (see for example the film  in the supplementary materials of this paper).

\subsection{Effect of large strains}\label{LargeStrains}

In order to include the influence of large strains in the  foundation-based approach, the full analytical description of the bond stress distribution was abandoned in works following Kaelble. The most important contribution is certainly that of Gent and Petrich  \cite{Gent1969} (hereafter named GP), which oversimplifies the mechanical description of the bonded region by assuming an inextensible and infinitely flexible tape backing. The adherence energy is thus estimated as the work of debonding in a way similar to eq.\ (\ref{G}):
\begin{equation}
\Gamma=a\int_0^{\varepsilon(\sigma_{c})}\sigma(\varepsilon,\dot\varepsilon) d\varepsilon,
\label{gammaGP}
\end{equation}
where the integral term represents the work per unit  volume to stretch the  fibrils up to the debonding stress $\sigma_c$.\footnote[2]{We remark that alternative debonding criteria have been proposed in the literature, as discussed in detail in.\cite{Yarusso1999} This does not affect our arguments, as long as these criteria are based on intensive quantities such as a maximum strain or an elastic energy density.} Since this work is entirely lost at fibril debonding, the adherence energy is dissipated through viscous friction during the adhesive stretching and through elastic hysteresis when fibrils debond. In this model also, such as in Kaelble's model, dissipation can occur even if the adhesive is purely elastic, in which case the integral in eq.\ (\ref{gammaGP}) is a density of stored mechanical energy, as the $\mathcal{W}$ term in Kaelble's model.
But for a viscoelastic material, since $\sigma(\varepsilon,\dot\varepsilon)$ represents the more general and non-linear response of the adhesive, including in particular viscous losses, the integral does not simply correspond to a density of stored mechanical energy.

\begin{figure}[t]
\centering
\includegraphics[width=8cm]{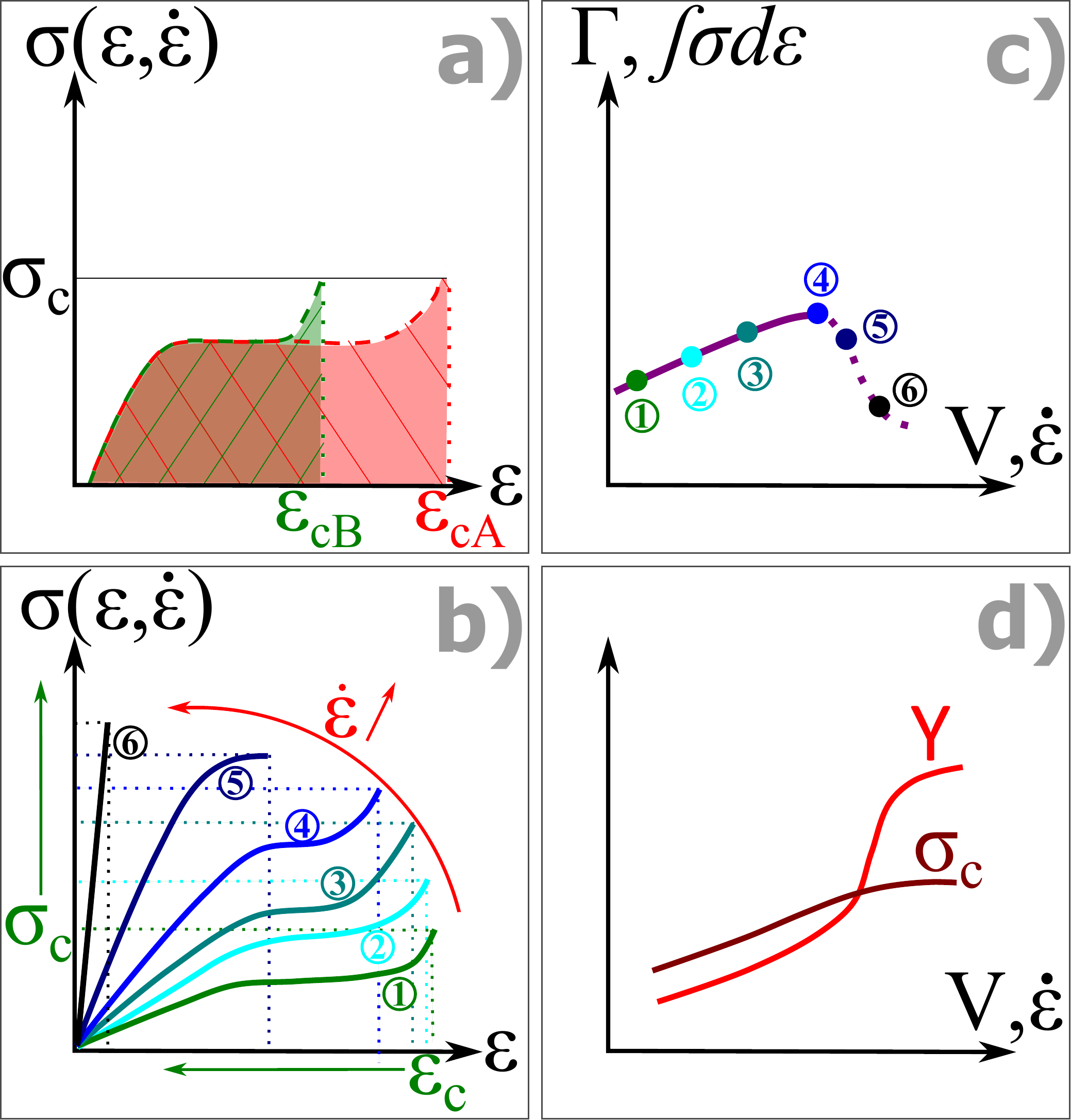}
 \caption{Interpretation of the shape of the $\Gamma(V)$ curve according to GP (a, b and c) and Kaelble's  (d) models.  (a): Work per unit volume to stretch a fibril up to debonding at a low peeling velocity, for two differently cross-linked adhesives with the same linear rheology.  (b): Schematic behaviour of a polymer fibril under elongation at different strain rates.  (c): Adherence energy estimated from the areas under the curves in picture (b), according to eq.\ (\ref{gammaGP}). (d): Evolution of the linear parameters used by Kaelble to interpret the $\Gamma(V)$ curve, such as the one in picture (c).  }\label{fig8}
\end{figure}

The most striking prediction of GP model is illustrated in Fig.\ \ref{fig8} (a): two adhesives with the same linear behaviour can possess very different adherence energies depending on the details of their non-linear behaviours. In particular, if we assume a simple criterion for debonding based on a critical stress $\sigma_c$, the adherence energy will increase if the fibrils can withstand larger stretches before debonding.

This effect can explain our observations on the increase of $\Gamma$ when the level of cross-linking decreases (see Fig.\ \ref{fig5}),  while keeping the linear rheology constant (in the relevant time-scales range): decreasing the level of cross-linking increases the deformation at which strain-hardening occurs and thus the maximum deformation that fibrils can sustain.  This experimental observation also proves that  models based on linear rheology cannot by definition predict the adherence energy of PSA.

However,  our data  show that the influence of cross-linking on $\Gamma$ is lost at high enough peeling velocities. This can also be understood using the principles of GP model in association with polymer mechanics arguments, because at such high velocities the network of entanglements does not have enough time to relax significantly. The elastic energy stored in the (dense) entanglement network will thus increase with the strain rate and become largely dominant compared to the energy stored in the (sparse) cross-linking network. The resulting stress build-up will cause the fibrils to debond  before  reaching the strain hardening domain at large deformations (see Fig.\ \ref{fig8} (b)) which is related to the  degree of cross-linking. The influence of the degree of cross-linking is therefore progressively lost when the peeling velocity increases and the behaviour becomes dominated by the entanglement network, which however still implies non-linearities in the rheology. Observations of Fig.\ \ref{fig6} (d) confirm this interpretation: fibrils do become shorter at higher peeling velocities, at least for the less cross-linked adhesives, and the lengths of the fibrils of adhesives of both cross-linking levels become comparable when $V$ increases.

The fact  the $\Gamma(V)$ curves with different cross-linking levels systematically collapse just before the peak in $\Gamma(V)$ (see Fig.\ \ref{fig5}) suggests that the decrease in fibril extensibility is also responsible for this peak.
Indeed, at the point where the effect of cross-linking disappears, the response of the adhesive is still in the non-linear regime and is dominated by the entanglement network. This condition corresponds to the transition between curves 4 and 5 in Fig.\ \ref{fig8} (b). A further increase in the strain rate would induce a decrease in $\Gamma$ (see Fig.\ \ref{fig8} (c)) as evaluated by eq.\ (\ref{gammaGP}), which corresponds to  a decrease of the area below the traction curve.

We finally note that, in the linear model of Kaelble (where $\Gamma\sim a\sigma_c^2/2Y$), this peak in $\Gamma(V)$ occurs when $Y$ increases faster than $\sigma_c^2$, at the onset of the glass transition,  as represented in Fig.\ \ref{fig8} (d). However, this explanation is not consistent with the high value of $\Gamma$ (which would imply linear deformations of several 100\%) and with the  large strains observed in Fig.\ \ref{fig6}. Once again, non-linearities and large strain behaviour are key elements to determine the $\Gamma(V)$ curves, even close to the peak in $\Gamma(V)$.

\section{Conclusion}

Our measurements and modelling considerations allow drawing the following conclusions regarding the peeling mechanics of PSA, and more generally of strongly confined soft viscoelastic materials:
\begin{itemize}
\item
The bond stress distribution inside the confined adhesive is essential to understand the dependence of the adherence energy $\Gamma$ on the geometry of loading, particularly on the adhesive thickness and on the peeling angle. We have demonstrated in this paper that $\Gamma$ increases with the peeling angle and that this dependence is separable from the peeling velocity dependence; moreover, this angular dependence can be explained using the analytical explicit solution we have derived from Kaelble's model.
\item
The occurrence of large deformations is essential to  explain, thanks to rate-dependent elastic hysteresis, the high values of the adherence energy $\Gamma$ of PSA. The large strain rheology of the adhesive must therefore be taken into account in any effort to quantitatively predict the adherence energy.
\item
The strong confinement of the soft incompressible adhesive is a key feature to reach these large deformations through cavitation and stringing and to develop hysteretic dissipation. However, this makes the link between the local cohesive response of the joint and the rheology of the adhesive very complex: a non-linear, yet homogeneous, description of the adhesive layer is not even enough, since its response cannot be simply described by the uniaxial behaviour of the bulk adhesive. The response in the region of the joint before cavitation should rather be described by an \oe dometric response that experiences progressive unconfinement. As for the modelling of the response of the fibrillated region, it requires a better understanding of the mechanism of cavitation and of the spatial organization of the foam-like fibril structure. The bulk parameters used in the models described in this paper (Kaelble and GP) can therefore only be interpreted as effective parameters and cannot be easily linked to the classical rheological parameters. Probe-tack investigations of our model adhesives will certainly provide very interesting insights towards a sound comprehensive modelling of the adherence energy in peeling.
\end{itemize}

\section*{Acknowledgments}
We thank A. Aubertin, L. Auffray, B. Bresson, L. Olanier and R. Pidoux for their help in the conception of experiments, and K. Benson for synthesizing  the adhesives. We thank E. Barthel, A. Chateauminois, C. Fr\'etigny, J.-P. Hulin, G. Pallares and O. Ramos for fruitful discussions. This work has been supported by the French ANR through Grant \#12-BS09-014.

\footnotesize{
\bibliography{biblio} 
\bibliographystyle{rsc} 
}

\end{document}